# Entropy, Dynamics and Instantaneous Normal Modes in a Random Energy Model

T. Keyes

*Chemistry Department, Boston University, Boston MA 02215*

It is shown that the fraction, $f_u$, of imaginary-frequency instantaneous normal modes (INM) may be defined and calculated in a random energy model (REM) of liquids. The configurational entropy, $S_c$, and the averaged hopping rate among the states, R, are also obtained and related to $f_u$, with the results $R \sim f_u$ and $S_c = a + b*\ln(f_u)$. The proportionality between R and $f_u$ is the basis of existing INM theories of diffusion, so the REM further confirms their validity. A link to $S_c$ opens new avenues for introducing INM into dynamical theories. Liquid 'states' are usually defined by assigning a configuration to the minimum to which it will drain, but the REM naturally treats saddle-barriers on the same footing as minima, which may be a better mapping of the continuum of configurations to discrete states. Requirements for a detailed REM description of liquids are discussed.

## 1. Introduction

The instantaneous normal modes[1-3] (INM) are the eigenfunctions of the Hessian, the matrix of second derivatives of the potential energy U with respect to the mass-weighted atomic or molecular coordinates; the frequencies are the square roots of the eigenvalues. INM differ from conventional normal modes because they are obtained for finite-T configurations sampled from an equilibrium distribution. Consequently, in liquids or in finite-T solids, some INM have imaginary frequencies, corresponding to downward curvature of the U-surface. The fraction, $f_u$, of Im-$\omega$ is a measure of the time the system spends above the inflection points. We proposed[1] that $f_u$ should be proportional to the rate of barrier crossing in the configuration space, and thus to the self diffusion constant D. This picture is most appropriate to low-D states, e.g. supercooled liquids, and fits naturally with the[4] potential energy 'landscape' paradigm. Considerable effort[3,5-16] has gone into the INM approach to diffusion, culminating in[15,16] two recent papers. We showed[15] that the Im-$\omega$ of the molecular centers-of-mass accurately predict D over a range of ~3 decades for seven densities and eight temperatures of supercooled and near-melting $CS_2$. La Nave et. al. found[16] a similarly excellent result for water. Given the pitfalls that have been identified[3,6-15], the description is far more successful than one might expect. The obvious question, then, is why it works so well.

While prior work has been mostly based upon the association of Im-$\omega$ with barriers, we recently suggested[17] a connection with the configurational entropy, $S_c$. Almost simultaneously, La Nave et. al. demonstrated[16] a beautiful linear master plot of $S_c$ vs ln($f_{dw}$), where[9,12,14] $f_{dw}$ is the fraction of Im-$\omega$ modes with 'double well' potential energy profiles. The states included exhibit D~$f_{dw}$. The prospect of another route to physical properties is exciting. Adam and Gibbs suggested[18] that $S_c$ governs the slow relaxation in supercooled liquids. Despite considerable empirical success, there is no satisfactory derivation of the Adam-Gibbs relation. Perhaps this might be accomplished via INM.

The dynamics of supercooled liquids are so complex that all theories are approximate and INM is no exception. A simplified model would be very helpful, and here we turn to a random energy model[19] (REM). REM have played an important role[20] in protein folding, and we believe that they have much to contribute to liquids. In a landmark paper[19], Bryngelson and Wolynes (BW) consider a chain of N interacting amino acids of which N$\rho$ are in the native state; the fraction $\rho$ is the order parameter. The resulting hierarchical REM is not completely random, since the random properties are functions of $\rho$. For liquids we suggest, following Adam-Gibbs, that cooperative local regions and a favored local packing replace individual amino acids and the 'local' native state.

From the landscape point of view, states of a liquid are usually defined as the local minima, or *inherent structures*[21], of U. Following Stillinger and Weber[21], a configuration is assigned to the minimum to which it will drain. Dynamics is then naturally visualized as hopping among the minima via the saddle-barriers. While the barriers are higher-order critical points of the U-surface, the system is never considered as 'belonging' to them-they simply provide the pathways between the essential objects, the minima (zero order critical points). This may be deceptive, since a liquid configuration is likely to be closer, by any reasonable metric, to a barrier than to a minimum. Cavagna[22] has proposed a 'saddles ruled scenario' in which the system occupies, and hops among, critical points of any order. A major difficulty in pursuing this promising idea is that, while any configuration is easily assigned to a minimum, no unambiguous mapping to the full set of critical points has been given.

BW give the energy distribution $G(E,\rho)$ for states which, *a priori*, can be either minima or saddles. Critical points of all orders are on the same footing from the beginning. Of course the REM has no U-surface but BW define a minimum as a state for which all the connected states have higher energy. In an early INM paper[6] we briefly touched on the REM, suggesting that an nth order critical point should have n neighbors with lower energy, although other schemes might be possible. Thus $f_u$ may be calculated given the distribution, $G_c(E',\rho;E)$, of states connected to a state with energy E. In general the conditional $G_c$ is not the same as the unconditional G, but for simplicity BW assume that it is. The averaged rate R for hopping among the states naturally divides into contributions $R_A$ and $R_B$ from hops to higher (Cavagna mechanism A) and lower (mechanism B) energies; the division is already explicit in BW. Barriers are not defined in the REM. An activation energy equal to $\Delta E$ is associated with hops to higher energy states, while the activation energy for mechanism B is zero.

In this article we calculate $f_u$, $S_c$, and R in the REM, treating both saddles and minima as states. We thus obtain an explicit realization of the[22] 'saddles ruled scenario', and its relation to the Stillinger-Weber scheme is discussed. Different choices of states will lead to different definitions of R and of $S_c$. The result[16] of La Nave et. al., $S_c=a+b*\ln(f_u)$, is obtained and the physical basis of the connection between Im-$\omega$ and entropy is explained. Both $R_A$ and $R_B$ are shown to be $\sim f_u$, confirming the basis of INM diffusion theory. The assumption of uncorrelated neighbor energies is discussed. The REM appears to be an attractive model for testing INM theories and for liquids in general.

## 2. INM Calculations in a Random Energy Model

Transposing the BW model, we imagine a liquid composed of N cooperative local regions of which $N\rho$ are in the local 'ground state' of favored packing. The energy distribution of the states is

$$G(E,\rho) = [2\pi\Delta(\rho)]^{1/2}\exp\{-\frac{[E-\mathcal{E}(\rho)]^2}{2\Delta(\rho)^2}\}, \tag{1}$$

where $\mathcal{E}(\rho)$ is the average energy and $\Delta(\rho)$ is the width. While explicit expressions for the parameters will be required for many applications, here the information $\mathcal{E}(\rho) \sim \Delta(\rho)^2 \sim N$ is sufficient, where N is large. The total number of states with $\rho$ is denoted $\Omega(N\rho)$,

$$\Omega(N\rho) = \frac{N! \nu^{N(1-\rho)}}{[N(1-\rho)]!(N\rho)!}, \tag{2}$$

and $\nu$ is the number of 'excited states' for a local region.

With no U-surface, various anzatz are needed to provide a dynamical framework. Following BW, a state of the entire liquid is considered connected to neighbor states which differ by a change in a single local region; the number of neighbors is easily seen to be $N\nu$. The probabilities that a state chosen at random has energy less than or greater than E are denoted $p^<(E,\rho)$ and $p^>(E,\rho)$, and are given by the integral of $G(E',\rho)$ from - to E and from E to +, respectively; $p^< + p^> = 1$. The probabilities for connected states are $p_c^<(E,\rho)$ and $p_c^>(E,\rho)$, obtained by replacing $G(E',\rho)$ with $G_c(E',\rho;E)$ in the prescription above. Thus, using the BW rule the probability that a state is a local minimum is $p_{min} = (p_c^>)^{N\nu}$, where E,$\rho$ arguments will be suppressed whenever the meaning of the resulting expressions seems obvious.

We proposed[6] that an nth order critical point should have $N\nu-n$ neighbors with higher energy and n with lower energy, with probability $p_n = (N\nu)!/[(N\nu-n)!n!] (p^<)^n (p^>)^{N\nu-n}$. The fraction of 'Im-$\omega$' is then $n/N\nu$ and

$$f_u(E,\rho) = \sum_{n=0}^{N\nu} (n/N\nu) \frac{N!}{(N\nu-n)!n!} (p_c^<)^n (p_c^>)^{N\nu-n} = p_c^<(E,\rho). \tag{3}$$

Eq. 3 is the key to expressing physical quantities in the REM in terms of the fraction of Im-$\omega$ modes. Assuming uncorrelated neighbor energies and with the further argument[19] that $G(E,\rho) \sim G(E,\rho\pm(1/N))$ it follows that $p^< = p_c^<$, $p^> = p_c^>$ and $f_u = p^<$. We will use this simplification in some of the arguments below, but we will strive for greater generality whenever possible.

**Imaginary frequency modes and the configurational entropy**

Turning to thermodynamics, the partition function is

$$Q(T) = \sum_{N\rho=0}^{N} \int_0^\infty dE \, \Omega(N\rho) \, G(E,\rho) \exp(-E/T) \quad (4)$$

$$\equiv \exp(-E(T) + S_c(T)),$$

with units such that $k_B=1$, and functions of T only are ensemble averages. Eq 4 incorporates the nontrivial assumption that all REM states, minima and saddles, contribute equally. The entropy is identified as $S_c$ because the REM has no vibrations and all degrees of freedom are 'configurational'.

In the inherent structure scheme, every configuration, no matter how close to a barrier, is assigned to a minimum. Moving among saddle-barriers in the same basin does not change the configuration-it is highly anharmonic 'vibration'. Thus[16,23] $S_c$ is obtained as $S-S_{vib}$, where $S_{vib}$ is the vibrational entropy and contains both harmonic and anharmonic contributions for a system confined to a basin. This definitions of $S_c$ is not equivalent to that of Eq 4. It will be identical at low T, since BW show that essentially all states are minima below a crossover energy $E_c$. The fundamental definition of configurational entropy is $S_c(T)=S_{liq}(T)-S_{xtl}(T)$, the difference between the liquid and crystal entropies at the same T. We suggest that moving among the barriers connected to a single basin is a liquid-like feature absent in the crystal and should be included in $S_c(T)$. Again, we hope that at the low T of greatest interest any numerical difference between the two versions of $S_c$ is small.

Writing the entire summand-integrand of Eq 4 as an exponential, the exponent is O(N) and for a given $\rho$ will be dominated by a most probable E, denoted $E^*$. Expanding the exponent to second order about $E^*$ and performing the Gaussian integration,

$$Q(T) = \sum_{N\rho=0}^{N} \exp[-E^*(T,\rho)/T + S_c(T,\rho)], \quad (5a)$$

where

$$S_c(T,\rho) = \ln[\Omega(N\rho) G(E^*(T,\rho),\rho) \sqrt{2\pi} \, \Delta(\rho)], \quad (5b)$$

and

$$E^*(T,\rho) = \varepsilon(\rho) - \Delta(\rho)^2/T. \quad (5c)$$

Usually a 'thermodynamic' $\rho$ should dominate and $S_c(T) = S_c(T,\rho^*(T))$.

We now relate the entropy to the averaged Im-$\omega$ fraction, denoted $f_u(T)$,

$$f_u(T) = \frac{\sum_{N\rho=0}^{N} \Omega(N\rho)\, q(T,\rho)\, \langle f_u(T,\rho)\rangle}{\sum_{N\rho=0}^{N} \Omega(N\rho)\, q(T,\rho)}, \qquad (6)$$

where the average at fixed $\rho$ is

$$\langle f_u(T,\rho)\rangle = \int_{-\infty}^{\infty} dE\, G(E,\rho)\, \exp(-E/T)\, f_u(E,\rho)\, /\, q(T,\rho), \qquad (7)$$

and the constant-$\rho$ partition function is

$$q(T,\rho) = \int_{-\infty}^{\infty} dE\, G(E,\rho)\, \exp(-E/T)$$
$$= \exp[-E^*(T,\rho)/T - \Delta(\rho)^2/(2T^2)]. \qquad (8)$$

The second equality is obtained with a Gaussian approximation for the integrand. Eqs 6-8 may be used to average any quantity X by replacing $f_u(E,\rho)$ with $X(E,\rho)$.

According to Eq 5c $E^*$ lies $O(N)$ below the center of the distribution, which has width $O(\sqrt{N})$. Thus $p^<$ may be evaluated using the asymptotic expansion of the error function,

$$p^<(E,\rho) = \frac{\Delta(\rho)}{\sqrt{2\pi}\,(\mathcal{E}(\rho)-E)}\, \exp\{-\frac{[E-\mathcal{E}(\rho)]^2}{2\Delta(\rho)^2}\}. \qquad (9)$$

The essence of the connection between Im-$\omega$ and entropy is now visible. From Eq. 5b the T-dependence of $S_c$ is determined by $\ln[G(E^*,\rho^*)]$ and (Eqs 1 and 9) $G(E^*,\rho^*) \sim p^<(E^*,\rho^*)$. In the uncorrelated REM $\langle p^<\rangle = f_u$ and the average would ordinarily be determined by the dominant E and $\rho$, $p^<(E^*,\rho^*)=f_u$; thus $S_c=a+\ln(f_u)$, the result[16] of La Nave et. al. The physical reason for the relation is very simple. The fraction $f_u$ of directions with 'downward curvature' at $E^*$ is proportional, absent correlation, to the number of states with energy less than $E^*$, which is also roughly the number of states available to the system, which determines $S_c$.

However the situation is unusual if $p_c^< = p^<$. With the $N\nu$ neighbors of a state spread out over the full distribution, the exponent (Eq 9) is O(N). The function being averaged in Eq 7 has E-dependence as strong as that of the weighting factors, $G(E)\exp(-E/T)$, the maximum of the integrand is shifted from $E^*$, and

$$<f_u(T,\rho)> \; = \; <p^<(T,\rho)> \; = \; \frac{T\exp[-\frac{\Delta(\rho)^2}{4T^2}]}{\sqrt{\pi}\,\Delta(\rho)} \; . \tag{10}$$

Referring to Eqs 1 and 5c it is seen that the *square* of the RHS has the same strong (exponential) T-dependence as $G(E^*)$ and (Eq 5b),

$$S_c(T,\rho) \; = \; \ln\{\Omega(N\rho)[<f_u(T,\rho)>\sqrt{\pi}\,\Delta(\rho)/T]^2\}$$
$$= \; a \; + \; 2\ln(<f_u(T,\rho)>) \; . \tag{11}$$

With dominance of a single $\rho^*$ a linear relation between $S_c$ and $\ln(f_u)$ holds again.

Nonetheless, we expect that $S_c = a + \ln(f_u)$ is correct for liquids. With no correlation almost all the neighbors of a state are within O($\sqrt{N}$) of $\mathcal{E}(\rho)$, while a thermally significant state has $E \sim E^*$, O(N) below $\mathcal{E}(\rho)$. Thus there are essentially no lower-energy neighbors and $f_u \sim \exp(-N)$. This is not correct for liquids although it is essential for Eq 11, where $\ln(f_u)$ must be O(N). As a simple alternative example, suppose the connected distribution is obtained from $G(E,\rho)$ by increasing the width so that the neighbor energy differences E'-E are O(1), $G_c(E',\rho;E) \sim \exp[-(E'-\mathcal{E}(\rho))^2/(2N\Delta(\rho)^2)]$. Then $f_u(E) \sim p_c^<(E) \sim G(E)^{1/N}$, $f_u$ *is* dominated by $E^*$, $G(E^*) \sim f_u(E^*)^N$, and $S_c/N = c' + \ln(f_u)$, with both $\ln(f_u)$ and c' O(1). These are the correct N-dependences for liquids; the relation between $S_c$ and $\ln(f_u)$ is robust.

**Imaginary frequency modes and the hopping rate**

The unaveraged hopping rate $R(E,\rho)$ is given in Eq 121 of BW, already divided into two terms corresponding to Cavagna's mechanism A and B. The escape rate from a state with E to one with E' is given by $R_0 \exp(-E_A/T)$, with $E_A = (E'-E)$ for E'>E (mechanism A) and $E_A = 0$ for E'<E (mechanism B). Recalling that there are $N\nu$ neighbors, averaging for fixed $\rho$ yields

$$<R_A(T,\rho)> = R_0 N\nu \int_{-\infty}^{\infty} dE\, G(E,\rho)\exp(-E/T)$$

$$* \int_{E}^{\infty} dE'\, \exp(-(E'-E)/T)\, G_c(E',\rho;E)\, /\, q(T,\rho) \qquad (12)$$

and

$$<R_B(T,\rho)> = R_0 N\nu \int_{-\infty}^{\infty} dE\, G(E,\rho)\exp(-E/T)\, p_c^<(E,\rho)\,/\,q(T,\rho)$$

$$= R_0 N\nu <f_u(T,\rho)> . \qquad (13)$$

The connection between $R_B$ and $f_u$ is *exact*, and does not require $G_c=G$.

We evaluate $<R_A(T,\rho)>$ by assuming $G_c=G$, and by dividing the E-integral into the contributions $R_A^1$ from - to $E^*$ and $R_A^2$ from $E^*$ to + . The factors of $\exp(-E/T)$ cancel and the E' integral is easily performed. The product $G(E')\exp(-E'/T)$ is sharply peaked at $E^*$, so for any $E<E^*$ the E' integral is just $q(T,\rho)$ and

$$<R_A^1(T,\rho)> = R_0 N\nu p^<(E^*,\rho) . \qquad (14)$$

As discussed above, $p^<(E^*,\rho)$ would usually equal $<p^<(T,\rho)>=<f_u(T,\rho)>$, but with $G_c=G$ the strong E dependence of $p^<$ leads (Eq 10) to a different result,

$$<R_A^1(T,\rho)> = R_0 N\nu (\sqrt{\pi}/2)(\Delta(\rho)/T)<f_u(T,\rho)>^2. \qquad (15)$$

We anticipate that use of a more reasonable $G_c$ would restore $R_A^1 \sim f_u$. For $E>E^*$ the asymptotic expansion of the E' integral and some algebra yields $<R_A^2>=<R_B>$, and

$$<R_A(T,\rho)> = R_0 N\nu <f_u(T,\rho)> [1 + (\sqrt{\pi}/2)(\Delta(\rho)/T)<f_u(T,\rho)>] . \qquad (16)$$

For the uncorrelated model the second term in the bracket is negligible ($f\sim\exp(-N)$) and, despite the behavior of $R_A^1$, $R_A \sim R_B \sim R \sim f_u$. The dominant contribution to $R_A$ comes from hops to states with $\mathcal{E}(\rho)>E'>E^*$, which is reasonable since essentially all statistically significant states lie in that range. With a dominant $\rho$, the final result is that $R(T)$ is indeed proportional to $f_u$. Our arguments of an essential link between the Im-$\omega$ modes and the hopping rate are confirmed within the REM.

## 3. Discussion

The empirical evidence for proportionality between D and $f_u$ in liquids is[15,16] now very strong. The complexity of the U-surface, however, renders a theoretical proof of this relation impossible. Some Im-$\omega$ in liquids unquestionably correspond to non-diffusive anharmonicities and these must not be used to express D. Our study of $CS_2$ uses[15] center-of-mass modes to remove rotational anharmonicities, while la Nave et. al. use[16] only modes with double well U-profiles. Thus the REM, a simplified model which allows unambiguous INM calculations and still preserves some essential dynamics and statics, is most appealing. In the REM the hopping rate among the critical points is clearly proportional to $f_u$. Despite the theoretical challenges which arise for real liquids, the calculations just presented, along with the recent simulations[15,16] provide the strongest arguments yet of a fundamental connection between Im-$\omega$ and diffusion.

The REM has also allowed us to derive the result[16] of La Nave et. al., a linear relation between $S_c$ and $\ln(f_u)$. The physical basis of the relation is very simple: the fewer states below the thermodynamic $E^*$, the lower the configurational entropy and the fewer the number of directions with downward curvature. INM may now provide a new way to understand the role of $S_c$ in dynamics, suggested[18] by Adam and Gibbs but never proven satisfactorily. A newer proposal[24] is that of Dzugatov, $D^* \sim \exp(S_2)$, where $D^*$ is a scaled D and $S_2$ is the 'pair correlation entropy'; if $D \sim f_u$ and $S_c = a + \ln(f_u)$ we obtain the Dzugatov form, with $S_2 \rightarrow S_c$. Sciortino et. al. argue[12,16] that $f_u$ should vanish at the temperature $T_c$ where activated dynamics first becomes important. With all the critical points as states, this should correspond to the thermodynamic state being a minimum. Substituting Eq 3 into Eq of BW yields the probability that the system is in a minimum

$$P_{LM} = \exp[-N\upsilon f_u],$$

and nonzero $P_{LM}$, or activated dynamics, indeed requires $f_u \sim O(1/N)$.

At some steps in this first paper we have employed the uncorrelated approximation to the energy distribution of neighbor states, $G_c(E',\rho;E) = G(E,\rho)$, which yields incorrect N-dependences. However the important result is the interrelations between R and $f_u$ and $S_c$ and $f_u$, not the N-dependences of these quantities *per se*. Furthermore we have argued that the important $f_u$-dependences will hold up for a broad range of possible choices for $G_c(E',\rho;E)$, including those appropriate for liquids. Eq 13 for $R_B$ is exact and independent of the form of $G_c$, while $R_A$ was derived assuming $G_c = G$. Perhaps a $G_c$-independent exact result for $R_A$ might also exist, because both R and $f_u$ are governed by $G_c$. On the other hand $S_c$ is a functional of G only, so a relation to will depend on the form of $G_c$; nonetheless we believe any reasonable $G_c$ will give La Nave's result[16].

In the REM the system naturally moves about the critical points of all orders, with no special role for the minima. This contrasts with the usual procedure in liquids of assigning a configuration to the minimum to which it drains. Cavagna suggests[22] that the saddles should be treated explicitly in liquids, and the REM provides an easy way to do this. Although we indicated how to transpose the protein-REM to liquids, we never considered explicit values of the parameters or their ρ-dependences. This will be done in future work on modeling supercooled liquids with the REM. A better treatment of $G_c$ might also allow realization of Cavagna's hypothesis that mechanisms A and B have different T-dependence.

**ACKNOWLEDGMENTS**
This work was supported by NSF Grant CHE9708005. Discussions with Janamejaya Chowdhary, Emelia La Nave, Antonio Scala and Andrea Cavagna are gratefully acknowledged.

**REFERENCES**

1. G. Seeley and T. Keyes, J. Chem. Phys. **91**, 5581 (1989).

2. R. M. Stratt, Acc. Chem. Res. **28**, 201, (1995).

3. T. Keyes, J. Phys. Chem. A **101**, 2921 (1997).

4. F. Stillinger, *Science* **267**, 1935 (1995).

5. G. Seeley, B. Madan and T. Keyes, J. Chem. Phys. **95**, 3847 (1991).

6. B. Madan and T. Keyes, J. Chem. Phys. **98**, 3342 (1993).

7. P.Moore, J.Chem.Phys. **100**, 6709 (1994).

8. M. Cho, G. R. Fleming, S. Saito, I. Ohmine and R. M. Stratt, J. Chem. Phys. **100**, 6672 (1994).

9. S. Bembenek and B. B. Laird, Phys. Rev. Lett. **74**, 936 (1995); J. Chem. Phys. **104**, 5199 (1996)

10. T. Keyes, G. Vijayadamodar and U. Zurcher, J. Chem. Phys. **106**, 4651 (1997).

11. Wu-Xiong Li and T. Keyes, J. Chem. Phys. **107**, 7275 (1997).

12. F. Sciortino and P. Tartaglia, Phys. Rev. Lett. **78**, 2385 (1977).

13. J. Gezelter, E. Rabanai and B. Berne, J. Chem. Phys. **107**, 4618 (1997).

14. Wu-Xiong Li, T. Keyes, and F. Sciortino, J. Chem. Phys. **108**, 252 (1998).

15. Wu-Xiong Li and T. Keyes, J. Chem. Phys. **111**, 5503 (1999).

16. E. La Nave, A. Scala, F. Starr, F. Sciortino and H. E. Stanley, Phys. Rev. Lett, submitted (2000).

17. U. Zurcher and T. Keyes, Phys. Rev. E **60**, 2065 (1999).

18. G. Adam and J. H. Gibbs, J. Chem. Phys. **43**, 139 (1965).

19. B. Derrida, Phys. Rev. **B 24**, 2613 (1981).

20. J. Bryngleson and P. G. Wolynes, J. Phys. Chem. **93**, 6902 (1989); J. Onuchic, Z. Luthey-Schulten and P. G. Wolynes, Ann. Rev. Phys. Chem. **48**, 539 (1997).

21. F. Stillinger and T. A. Weber, Phys. Rev. **A28**, 2408 (1983).

22. A. Cavagna, Phys. Rev. Lett, submitted (2000), e-print cond-mat/9910244 (1999).

23. F. Sciortino, W. Kob and P. Tartaglia, Phys. Rev. Lett. **83**, 3214 (1999); B. Coluzzi, M. Mezard, J. Parisi and P. Verrocchio, J. Chem. Phys. **111**, 9039 (1999).

24. M. Dzugatov, Nature **381**, 137 (1996).